\documentstyle[preprint,aps,epsfig]{revtex}

\begin{document}
\draft
\preprint{}
\def\qslash{\hbox{q\kern-.5em\lower.1ex\hbox{/}}}


\begin{flushright}
USM-TH-79
\end{flushright}

\bigskip\bigskip
\centerline{\large \bf
The nucleon spin structure in a simple quark model}

\vspace{22pt}

\centerline{\bf Bo-Qiang Ma$^{a}$,  Ivan Schmidt$^{b}$, and
Jian-Jun Yang$^{b,c}$ }

\vspace{8pt} {\centerline {$^{a}$Department of Physics, Peking
University, Beijing 100871, China\footnote{Mailing address},}}

{\centerline {CCAST (World Laboratory), P.O.~Box 8730, Beijing
100080, China,}}

{\centerline {and Institute of Theoretical Physics, Academia
Sinica, Beijing 100080, China}

{\centerline{Email: mabq@phy.pku.edu.cn}}

{\centerline {$^{b}$Departamento de F\'\i sica,
Universidad T\'ecnica Federico Santa Mar\'\i a,}}

{\centerline {Casilla 110-V, 
Valpara\'\i so, Chile}}

{\centerline{Email: ischmidt@fis.utfsm.cl} }

{\centerline {$^{c}$Department of Physics, Nanjing Normal
University,}}

{\centerline {Nanjing 210097, China}}

{\centerline{Email: jjyang@fis.utfsm.cl} }


\begin{abstract}
We investigate the spin structure of the nucleon in an extended
Jaffe-Lipkin quark model. In addition to the conventional $3q$
structure, different $(3q)(Q\bar{Q})$  admixtures in the nucleon
wavefunction are also taken into account. The contributions to the
nucleon spin from various components of the nucleon wavefunction
are discussed. The effect due to the Melosh-Wigner rotation is
also studied.
\end{abstract}

{\centerline{PACS numbers: 12.39.-x, 13.60.Hb, 13.88.+e,
14.20.Dh}}


\vfill

\newpage

\section{Introduction}

It has been more than a decade since the discovery of the
Gourdin-Ellis-Jaffe sum rule (GEJ) \cite {GEJ} violation in the
polarized deep inelastic scattering (DIS) experiment by the
European Muon Collaboration \cite{EMC}. The physics community was
puzzled since the experimental data meant a surprisingly small
contribution to the proton spin from the spins of the quarks, in
contrast to the Gell-Mann-Zweig quark model in which the spin of
the proton is totally provided by the spins of the three valence
quarks. This gave rise to the proton  ``spin crisis" or ``spin
puzzle", and triggered a vast number of theoretical and
experimental investigations on the spin structure of the nucleon.
Among them, there was an interesting contribution to understand
the spin of the nucleon within a ``minimal" simple quark model
\cite{JL}, where it was observed that the nucleon has only a small
amplitude to be a bare three quark state $\left|qqq\right>$, while
the largest term in the wavefunction is $\left|qqqQ\bar Q\right>$,
in which $Q \bar Q$ denotes sea quark-antiquark pairs.

There was a prevailing impression that the proton spin structure
is in conflict with the quark model. However, there has been an
attempt to understand the proton spin puzzle within the quark
model by using the Melosh-Wigner rotation effect
\cite{Ma91b,Ma98a}, which comes from the relativistic effect of
the quark intrinsic transversal motion inside the proton. It was
pointed out \cite{Ma91b,Ma98a,Bro94} that the quark helicity
($\Delta q$) observed in polarized DIS is actually the quark spin
defined in the light-cone formalism, and it is different from the
quark spin ($\Delta q_{QM}$) as defined in the quark model. Thus
the small quark helicity sum observed in polarized DIS is not
necessarily in contradiction with the quark model in which the
proton spin is provided by the valence quarks \cite{Ma98a,Ma96}.
Recent progress \cite{Sch97,Ma98} has also been made on the
Melosh-Wigner rotation effect in other physical quantities related
to the spin structure of the nucleon, and the significance of the
Melosh-Wigner rotation connecting the spin states between the
light-front dynamics and the conventional instant-form dynamics
has been widely accepted. Thus it is necessary to check what 
can obtained for the spin structure of the nucleon within the 
quark model, after we take into account the Melosh-Wigner 
rotation. Certainly our present understanding of the nucleon spin
structure has been enriched from what we knew before the 
discovery of the GEJ sum rule violation, and we now know that 
both the sea quarks and the gluons play an important role in the 
spin structure of the nucleon. The purpose of this paper is to 
extend the simple Jaffe-Lipkin quark model to a more general 
framework, by including other necessary ingredients in the 
nucleon sea such as pseudoscalar mesons, whose 
addition is supported by available theoretical and experimental 
studies.

The paper is organized as follows. In Section II, we briefly
review the Melosh-Wigner rotation effect in the quark model, and
show that the introduction of an up($u$)-down($d$) quark flavor
asymmetry in the Melosh-Wigner rotation factors can reproduce the
present experimental data of the integrated spin structure
functions for both the proton and the neutron
\cite{SMC,E142,E143}, within a simple SU(6) quark model with only
three valence quarks. In Section III, we introduce the
contribution from the higher Fock states $\left| BM
\right>=\left|qqq Q\bar Q \right>$ in which the quark and
antiquark of an quark-antiquark pair are rearranged
non-perturbatively with the three valence quarks into a
pseudoscalar meson and a baryon, and we write the configuration as
a baryon-meson (BM) fluctuation \cite{BM}. It is shown that the
consideration of the lowest $p(uud D\bar D)=n(udD)\pi^+(u \bar D)$
fluctuation, which is supported by the observed Gottfried sum rule
violation \cite{NMC,GSRex}, introduces an $u$-$d$ flavor
asymmetric term in the quark contributions to the nucleon and
produces a reasonable $u$-$d$ Melosh-Wigner rotation asymmetry
which is required to reproduce the data. In this section, we point
out that the Jaffe-Lipkin term of quark-antiquark pairs (which are
actually vector mesons in a baryon-meson fluctuation picture) will
be important when there is need for negatively polarized
antiquarks. Thus we
present a new ``miminal" quark model extension of Jaffe-Lipkin
model, with three valence quarks, sea quark-antiquark pairs in
terms of baryon-meson fluctuations where the mesons are either
pseudoscalar or vector mesons, in order to
understand the nucleon spin structure within the quark model
framework. Finally, we present discussions and conclusions in
Section IV.

\section{The naive quark model and the Melosh-Winger rotation}

The spin-dependent structure functions for the proton and
the neutron, when expressed in terms of the quark helicity
distributions $\Delta q(x)$, should read
\begin{equation}
g_1^p(x)=\frac{1}{2}[\frac{4}{9}(\Delta u(x) +\Delta \bar u(x))
+\frac{1}{9}(\Delta d(x) +\Delta \bar d(x))
+\frac{1}{9}(\Delta s(x) +\Delta \bar s(x))],
\label{ejsp}
\end{equation}

\begin{equation}
g_1^n(x)=\frac{1}{2}[\frac{1}{9}(\Delta u(x) +\Delta \bar u(x))
+\frac{4}{9}(\Delta d(x) +\Delta \bar d(x))
+\frac{1}{9}(\Delta s(x) +\Delta \bar s(x))],
\label{ejsn}
\end{equation}
where the quantity $\Delta q(x)$
is defined by the axial current matrix element
\begin{equation}
\Delta q=\left <p,\uparrow \right|\overline{q} \gamma^{+}
\gamma_{5} q \left |p,\uparrow \right >.
\end{equation}
By expressing the quark axial charge or the quark helicity related
to the axial quark current $\bar q \gamma^{\mu} \gamma^5 q$ by
$\Delta Q=\int_0^1 {\rm d} x [\Delta q(x)+\Delta \bar q(x)]$, we
obtain
\begin{equation}
\Gamma^p=\int_0^1 {\rm d} x g_1^p(x)
=\frac{1}{2}(\frac{4}{9}\Delta U
+\frac{1}{9}\Delta D
+\frac{1}{9}\Delta S),
\end{equation}

\begin{equation}
\Gamma^n=\int_0^1 {\rm d} x g_1^n(x)
=\frac{1}{2}(\frac{1}{9}\Delta U
+\frac{4}{9}\Delta D
+\frac{1}{9}\Delta S).
\end{equation}
Two linear combinations of the axial charges, $\Delta Q^3=\Delta U
- \Delta D$ and $\Delta Q^8=\Delta U + \Delta D - 2 \Delta S$, are
therefore given by
\begin{equation}
\Delta Q^3=6 (\Gamma^p -\Gamma^n) =\Delta U - \Delta
D=G_A/G_V=1.261, \label{bsr}
\end{equation}
from neutron decay plus isospin symmetry, and by

\begin{equation}
\Delta Q^8=\Delta U + \Delta D - 2 \Delta S=0.675
\label{su3}
\end{equation}
from strangeness-changing hyperon decays plus flavor SU(3)
symmetry. Prior to the EMC experiment, the flavor singlet axial
charge was evaluated, by Gourdin and Ellis-Jaffe \cite{GEJ},
assuming $\Delta S=0$, to be

\begin{equation}
\Delta Q^0=\Sigma= \Delta U + \Delta D + \Delta S=\Delta Q^8.
\end{equation}
Then one obtains, neglecting small QCD corrections,
the GEJ sum rule,

\begin{equation}
\Gamma^p
=\frac{1}{12}\Delta Q^3
+\frac{1}{36}\Delta Q^8
+\frac{1}{9}\Delta Q^0=0.198,
\end{equation}
which is significantly larger than the observed experimental
result of 0.126 from the EMC experiment \cite{EMC}, but now
revised to be 0.136 \cite{SMC,E142,E143}.

The discovery of the GEJ sum rule violation came as a  big
surprise to the physics community since the sum of the quark
helicities $\Sigma$ inferred from Eqs.~(\ref{bsr}) and (\ref{su3})
and the observed $\Gamma^p$, by allowing $\Delta S \neq  0$, gave
the value

\begin{equation}
\Sigma=\Delta U + \Delta D + \Delta S=0.020
\end{equation}
from the EMC data $\Gamma^p=0.126$ \cite{EMC}, and
\begin{equation}
\Sigma=\Delta U + \Delta D + \Delta S \approx 0.30
\end{equation}
from the revised results $\Gamma^p=0.136$ and $\Gamma^n=-0.03$,
assuming SU(3) symmetry \cite{SMC,E142,E143}. This is in conflict
with the naive expectation that the spin of the proton is totally
provided by the spins of the three valence quarks in the naive
SU(6) quark model, if one interpreted the quark helicity
$\Delta Q$ as the quark spin contribution to the proton spin. Many
theoretical and experimental investigations have been devoted to
understanding this proton ``spin puzzle" or ``spin crisis"
\cite{Spin}.

However, it has been pointed in Refs.~\cite{Ma91b,Ma98a} that this
puzzle can be easily explained within the naive SU(6) quark model
if one properly considers the fact that the observed quark
helicity $\Delta Q$ is the quark spin defined in the light-cone
formalism (infinite momentum frame), and it is different from the
quark spin as defined in the rest frame of the nucleon (or in the
quark model). In the light-cone or quark-parton descriptions,
$\Delta q (x)=q^{\uparrow}(x)-q^{\downarrow}(x)$, where
$q^{\uparrow}(x)$ and $q^{\downarrow}(x)$ are the probabilities of
finding a quark or antiquark with longitudinal momentum fraction
$x$ and polarization parallel or anti-parallel to the proton
helicity in the infinite momentum frame. However, in the nucleon
rest frame one finds \cite{Ma91b,Bro94},
\begin{equation}
\Delta q (x) =\int [{\mathrm d}^2{\mathbf k}_{\perp}]
M_q(x,{\mathbf k}_{\perp}) \Delta q_{QM} (x,{\mathbf k}_{\perp}),
\label{Melosh1}
\end{equation}
with
\begin{equation}
M_q(x,{\mathbf k}_{\perp})=\frac{(k^+ +m)^2-{\mathbf k}^2_{\perp}}
{(k^+ +m)^2+{\mathbf k}^2_{\perp}},
\label{eqM1}
\end{equation}
where $M_q(x,{\mathbf k}_{\perp})$ is the contribution from the
relativistic effect due to the quark transverse motion (or the
Melosh-Wigner rotation effect), $q_{s_z=\frac{1}{2}}(x,{\mathbf
k}_{\perp})$ and $q_{s_z=-\frac{1}{2}}(x,{\mathbf k}_{\perp})$ are
the probabilities of finding a quark and antiquark with rest mass
$m$ and transverse momentum ${\mathbf k}_{\perp}$ and with spin
parallel and anti-parallel to the rest proton spin, $\Delta q_{QM}
(x,{\mathbf k}_{\perp})= q_{s_z=\frac{1}{2}}(x,{\mathbf
k}_{\perp})- q_{s_z=-\frac{1}{2}}(x,{\mathbf k}_{\perp})$, and
$k^+=x {\cal M}$, where ${\cal M}^2=\sum_{i}\frac{m^2_i+{\mathbf
k}^2_{i \perp}}{x_i}$. The Melosh-Wigner rotation factor
$M_q(x,{\mathbf k}_{\perp})$ ranges from 0 to 1; thus $\Delta q$
measured in polarized deep inelastic scattering cannot be
identified with $\Delta q_{QM}$, the spin carried by each quark
flavor in the proton rest frame or the quark spin in the quark
model.

We now check whether it is possible to explain the observed data
for $\Gamma^p$ and $\Gamma^n$ within the SU(6) naive quark model
by taking into account the Melosh-Wigner rotation effect. Though
we do not expect this to be the real situation, it is interesting
since there existed a general impression that it is impossible to
explain the proton ``spin puzzle" within the SU(6) naive quark
model. Also an early attempt \cite{Ma91b} for such purpose
failed, by using the early EMC data $\Gamma^p=0.126$ and
$\Gamma^n$ obtained from the Bjorken sum rule
$\Gamma^p-\Gamma^n=\frac{1}{6}G_A/G_V$. We start from the
conventional SU(6) naive quark model wavefunctions for the proton
and the neutron
\begin{equation}
|p ^{\uparrow} \rangle =\frac{1}{\sqrt{18}}( 2|u ^{\uparrow} u
^{\uparrow} d ^{\downarrow}  \rangle -|u ^{\uparrow}
u^{\downarrow} d^{\uparrow}  \rangle -|u^{\downarrow} u^{\uparrow}
d^{\uparrow}  \rangle) +(cyclic \hspace{0.2cm} permutation);
\end{equation}
\begin{equation}
|n ^{\uparrow} \rangle =\frac{1}{\sqrt{18}}( 2|d^{\uparrow}
d^{\uparrow} u^{\downarrow}  \rangle -|d^{\uparrow} d^{\downarrow}
u^{\uparrow}  \rangle -|d^{\downarrow} d^{\uparrow} u^{\uparrow}
\rangle) +(cyclic \hspace{0.2cm} permutation).
\end{equation}
One finds that the quark spin contributions $\Delta
u_{QM}=\frac{4}{3}$, $\Delta d_{QM}=-\frac{1}{3}$, and $\Delta
s_{QM}=0$ for the proton, and  the exchange of $u \leftrightarrow
d$ in the above quark spin contributions gives those for the
neutron. Then we get the integrated spin structure functions for
the proton and the neutron as
\begin{equation}
\Gamma^p=\frac{1}{2}(\frac{4}{9} \langle M_u \rangle
\Delta u_{QM}
+\frac{1}{9} \langle M_d \rangle \Delta d_{QM}
+\frac{1}{9} \langle M_s \rangle \Delta s_{QM});
\label{gpqm}
\end{equation}
\begin{equation}
\Gamma^n=\frac{1}{2}(\frac{1}{9} \langle M_u \rangle
\Delta u_{QM}
+\frac{4}{9} \langle M_d \rangle \Delta d_{QM}
+\frac{1}{9} \langle M_s \rangle \Delta s_{QM}),
\label{gnqm}
\end{equation}
where $\langle M_q \rangle$ is the averaged Melosh-Wigner rotation
factor for the quark $q$. From Eqs.~(\ref{gpqm}) and (\ref{gnqm})
we obtain
\begin{equation}
\langle M_u \rangle \Delta
u_{QM}=\frac{24\Gamma^p-6\Gamma^n-\langle M_s\rangle \Delta s
_{QM}}{5};
\end{equation}
\begin{equation}
\langle M_d \rangle \Delta
d_{QM}=\frac{24\Gamma^n-6\Gamma^p-\langle M_s\rangle \Delta s
_{QM}}{5},
\end{equation}
from which we get the values
\begin{equation}
\langle M_u \rangle \Delta u_{QM}=0.689 \label{vMu};
\end{equation}
\begin{equation}
\langle M_d \rangle \Delta d_{QM}=-0.307 \label{vMd},
\end{equation}
with the inputs $\Gamma^p=0.136$, $\Gamma^n=-0.03$
\cite{SMC,E142,E143}, and $\Delta s_{QM}=0$. Thus we get, for
$\Delta u_{QM}=\frac{4}{3}$ and $\Delta d_{QM}=-\frac{1}{3}$, that
\begin{equation}
\langle M_u \rangle=0.517, ~~\langle M_d \rangle=0.921,
~~~{\mathrm and}~~~ r_{d/u}=\langle M_d \rangle/\langle M_u
\rangle=1.78, \label{Mud}
\end{equation}
which means that we need a flavor asymmetry between the $u$ and
$d$ quarks for the Melosh-Wigner rotation factors to reproduce the
observed data $\Gamma^p$ and $\Gamma^n$ within the SU(6) naive
quark model. The sum of quark helicities in this situation is
\begin{equation}
\Sigma=\langle M_u \rangle \Delta u_{QM}+\langle M_d \rangle
\Delta d_{QM}+\langle M_s \rangle \Delta s_{QM} \approx 0.38,
\end{equation}
which is small and far from 1, the total quark spin contributions
$\Delta u_{QM}+\Delta d_{QM} + \Delta s_{QM}$ to the nucleon spin.
We need to point out here that there is no mistake in calling the
quark helicity $\Delta q =\langle M_u \rangle \Delta u_{QM}$ the
quark spin contribution as commonly accepted in the literature, if
one properly understands it from a relativistic viewpoint. But in
this case there should be also non-zero contribution to the {\it
relativistic} orbital angular momentum even for the S-wave quarks
in the naive SU(6) quark model. Detail illustrations concerning
this point can be found in Ref.~\cite{Ma98} where the role played
by the Melosh-Wigner rotation on the quark orbital angular
momentum is studied.

We know that a symmetry between the valence $u(x)$ and $d(x)$
quark distributions would mean $F^n_2(x)/F^p_2(x) \geq
\frac{2}{3}$ for the unpolarized structure functions $F_2(x)$ in
the whole $x$ region $x=0 \to 1$, and this has been ruled out by
the experimental observation that $F^n_2(x)/F^p_2(x) < 0.5$ at $x
\to 1$. This indicates an asymmetry between the $u(x)$ and $d(x)$
valence quark distributions, and such an asymmetry, which can be
reproduced in an SU(6) quark-spectator-diquark model
\cite{Fey72,DQM}, also implies an asymmetry between the
Melosh-Wigner rotation factors for $\langle M_u \rangle$ and
$\langle M_d \rangle$ \cite{Ma96}. It is interesting to notice
that the asymmetry ratio $r_{d/u}=\langle M_d \rangle / \langle
M_u \rangle$ larger than 1 is in the right direction as predicted
in the quark-spectator-diquark model \cite{Ma96}, though the
magnitude is not so big as that given in Eq.~(\ref{Mud}). This may
imply that an additional source for a bigger $u-d$ flavor
asymmetry is needed for a more realistic description of the
nucleon.

\section{The intrinsic nucleon sea from the baryon-meson fluctuations}

Though the proton spin problem raised doubt about the quark model
at first, there has been a consistent attempt to understand the
problem within the quark model framework on extended quark models
\cite{JL,Keppler95,Qing98}, and also on the quark model in the
light-cone formalism \cite{Ma91b,Ma98a,Bro94,Ma96,Sch97}. For
example, Jaffe and Lipkin \cite{JL} found that both the EMC data
and the $\beta$-decay data can be fitted using a ``reasonable
modification" of the standard quark model in which the only
additional degrees of freedom are a single quark-antiquark pair in
the lowest states of spin and orbital motion allowed by
conservation laws. Keppler {\it et al.} \cite{Keppler95} pointed
out that the $5q$ component should be dominated by pseudoscalar
S-wave mesons. Qing, Chen, and Wang \cite{Qing98} gave a numerical
calculation of the coefficients of the total wavefunction in the
non-relativistic quark potential model by including the
Melosh-Wigner rotation effect \cite{Ma91b}, although in a
different manner, and showed that the proton wavefunction is
dominated by the bare $3q$ state.

In this section, we will perform a more detailed analysis of the
spin structure in an extended quark model by taking into account
the higher Fock states in the wavefunction of the proton, and
check how these higher Fock states may influence the analysis in
Section II, where we considered the effect of the 
Melosh-Wigner rotation with the three valence quark component 
only. In the higher Fock states, the quark and antiquark of a 
quark-antiquark pair are rearranged non-perturbatively with the 
three valence quarks into a meson and a baryon and we write the 
configuration as a baryon-meson fluctuation. In the ``minimal" 
quark model of Jaffe-Lipkin \cite{JL}, the quark-antiquark pairs 
are actually vector mesons in a baryon-meson fluctuation picture.
The higher Fock state in the ``minimal" quark model, which is 
referred to as Jaffe-Lipkin term, can be written as

\begin{equation}
|[JL] ^\uparrow  \rangle= \rm{cos}\theta |[b\varepsilon]^\uparrow
\rangle +\rm{sin}\theta |[bD] ^\uparrow  \rangle, \label{JL}
\end{equation}
where $b$ denotes the three quark $qqq$ component for a bare
nucleon. The extra $qqq Q\bar{Q}$ component
$|[b\varepsilon]^\uparrow \rangle$ with the $0^{++}$ $Q\bar{Q}$
denoted by $\varepsilon$ can be written as,

\begin{equation}
|\varepsilon \rangle= \sqrt{\frac{1}{3}} |Y^\Uparrow X^\Downarrow
\rangle +\sqrt{\frac{1}{3}} |Y^\Downarrow X^\Uparrow \rangle
-\sqrt{\frac{1}{3}} |Y^0 X^0 \rangle,
\end{equation}
and the extra $qqq Q\bar{Q}$ component $|[bD]^\uparrow \rangle$
with the $1^{++}$ $Q\bar{Q}$ denoted by $D$ can be written as,

\begin{equation}
|[bD]^\uparrow \rangle= \sqrt{\frac{2}{3}} |b^\downarrow
D^\Uparrow \rangle -\sqrt{\frac{1}{3}} |b ^\uparrow D^0 \rangle,
\end{equation}
with
\begin{equation}
|D^\Uparrow \rangle= \sqrt{\frac{1}{2}} |Y^\Uparrow X^0\rangle
-\sqrt{\frac{1}{2}} |Y^0 X^\Uparrow \rangle;
\end{equation}

\begin{equation}
|D^0 \rangle= \sqrt{\frac{1}{2}} |Y^\Uparrow X ^\Downarrow\rangle
-\sqrt{\frac{1}{2}} |Y^\Downarrow X^\Uparrow \rangle,
\end{equation}
where $D^\Uparrow$, $D^0$, and $D^\Downarrow$ denote the $J_3$
states of the $Q\bar Q$ pair; $Y^\Uparrow$, $Y^0$, and
$Y^\Downarrow$ denote the $L_3$ states of the $Q\bar Q$ spin;
$X^\Uparrow$, $X^0$, and $X^\Downarrow$ denote the $S_3$ states of
the $Q\bar Q$ spin; and $\Uparrow$ denotes a $J_3=1$ spin
contribution and $\uparrow$ denotes a $J_3=1/2$ spin contribution.
With the above higher Fock states included, Jaffe and Lipkin found
that the proton state has only a small amplitude to be a bare
three-quark baryon state, in order to reproduce the large negative
sea spin found in their analysis on the hyperon beta decay, baryon
magnetic moments and the EMC result on the fraction of the spin of
the nucleon carried by the spins of the quarks \cite{JL}.

In the Jaffe-Lipkin term, only P-wave vector $q\bar{q}$ pairs have
been taken into account. However, if we consider the $qqq Q\bar Q$
component as a baryon-meson fluctuation of the nucleon, then the
dominant fluctuations should be the ones in which the baryon-meson
has the smallest off-shell energy \cite{BM}. Therefore energy
considerations require that the $qqq Q\bar Q$ component should be
dominated by pseudoscalar S-wave mesons, like the pion
\cite{Keppler95}. In order to describe a nucleon state more
realistically, we include these new higher Fock states in addition
to the Jaffe-Lipkin states, and the nucleon state should be in
principle extended to

\begin{equation}
|B^\uparrow \rangle = \rm{cos}\alpha \rm{cos}\beta |b^\uparrow
\rangle +\rm{sin}\alpha \rm{cos}\beta |[BM] ^\uparrow  \rangle
+\rm{sin}\beta |[JL] ^\uparrow  \rangle \label{JLBM},
\end{equation}
where $\alpha$ and $\beta$ are the mixing angles between the bare
baryon state and the baryon-meson states $|[BM] ^\uparrow \rangle
$ and $|[JL]^\uparrow \rangle$, and the baryon-meson BM state can
be written as

\begin{equation}
|[BM] ^\uparrow \rangle = \sqrt{\frac{2}{3}} |b ^\downarrow M
Y^\Uparrow \rangle -\sqrt{\frac{1}{3}} |b ^\uparrow M Y^{0}
\rangle,
\label{BM}
\end{equation}
where $M$ denotes the spin contribution from the pseudoscalar
meson (with spin zero but parity -1), and $Y$ denotes orbital
angular momentum (with $L=1$) due to the relative motion between
the baryon and the meson. We can also extend the BM term by
including the $b^*=qqq$ state with spin $S=3/2$ if higher order
baryon-meson fluctuations need to be considered, and in this case
we write
\begin{equation}
|[BM] ^\uparrow \rangle = A(bM)|[bM] ^\uparrow \rangle
+A(b^*M)|[{b^*}M] ^\uparrow \rangle,
\end{equation}
where
\begin{equation}
|[bM] ^\uparrow \rangle = \sqrt{\frac{2}{3}} |b ^\downarrow M
Y^\Uparrow \rangle -\sqrt{\frac{1}{3}} |b ^\uparrow M Y^{0}
\rangle,
\end{equation}
as in Eq.~(\ref{BM}), and
\begin{equation}
|[{b^*}M] ^\uparrow \rangle = \sqrt{\frac{1}{2}}
|{b^*}^{\Uparrow\uparrow} M Y^\Downarrow \rangle
-\sqrt{\frac{1}{3}} |{b^*}^{\uparrow} M Y^{0} \rangle
+\sqrt{\frac{1}{6}} |{b^*}^{\downarrow} M Y^{\Uparrow} \rangle .
\end{equation}
The anti-quarks are unpolarized since they exit only in the
pseudoscalar meson of the BM state.

Using the wavefunction (\ref{JLBM}), we now calculate the
contributions $\Sigma_v$, $\Sigma_s$, and $\Lambda_s$, of the
valence quark spins, the spin of the sea, and the orbital angular
momentum of the sea, to the spin of the proton, and we obtain

\begin{equation}
\Sigma_v=\rm{cos}^2 \alpha \rm{cos}^2 \beta-\frac{1}{3}
\rm{sin}^2\alpha \rm{cos}^2\beta+\rm{sin}^2 \beta \rm{cos}^2
\theta -\frac{1}{3} \rm{sin}^2 \beta \rm{sin}^2 \theta; \label{Sv}
\end{equation}

\begin{equation}
\Sigma_s=\frac{8}{3}\sqrt{\frac{1}{2}}\rm{sin}^2 \beta \rm{sin}
\theta \rm{cos}\theta + \frac{2}{3} \rm{sin}^2\beta \rm{sin}^2
\theta; \label{Ss}
\end{equation}

\begin{equation}
\Lambda_s=-\frac{8}{3}\sqrt{\frac{1}{2}}\rm{sin}^2 \beta \rm{sin}
\theta \rm{cos}\theta + \frac{2}{3} \rm{sin}^2\beta \rm{sin}^2
\theta + \frac{4}{3} \rm{sin}^2\alpha \rm{cos}^2\beta,
\end{equation}
with
\begin{equation}
\Sigma_v+\Sigma_s+\Lambda_s=1.
\end{equation}
We can say alternatively that $\Sigma_v$ comes from the spin sums
of all $b=qqq$ terms, $\Sigma_s$ from the spin sums of all $Q\bar
Q$ terms ($X$ terms in the Jaffe-Lipkin term and $M$ terms in the
BM term Eq.~(\ref{BM})), and $\Lambda_s$ from the orbital angular
momentum of all $Y$ terms in the nucleon state $|B^\uparrow
\rangle$.

It can be easily seen that the sea spin $\Sigma_s$ comes entirely
from the Jaffe-Lipkin term, since the spin contribution from the
$M$ terms is zero. It is also interesting that $\Sigma_s$ cannot
be negative if there is no interference between the two components
$\left|b\varepsilon\right>$ and $\left|b D\right >$ in the
Jaffe-Lipkin term Eq.~(\ref{JL}). If we follow Ref.~\cite{JL} and
adopt the two models for the sea spin $\Sigma_s$, then we find
that we must arrive at the conclusion of Jaffe-Lipkin term
dominance. In the first model (called II in Ref.~\cite{JL}), the
sea is taken as $\rm{SU(3)_{flavor}}$ symmetric, and
$\Sigma_s(\rm{II})=-0.69\pm 0.27$. In the second model (called III
in Ref.~\cite{JL}), the sea is taken as $\rm{SU(2)_{flavor}}$
symmetric, and $\Sigma_s(\rm{III})=-0.56\pm 0.22$. On the other
hand, the data on hyperon and nucleon $\beta$-decays requires
$\Sigma_v$ to be approximately  $\frac{3}{4}$. Of course, it is
impossible for us to completely determinate $\alpha$, $\beta$ and
$\theta$ using the values of $\Sigma_v$ and $\Sigma_s$ mentioned
above. But, taking (\ref{Sv}) and (\ref{Ss}) as constraint
conditions, we can give a range of values of these mixing angles.
Selected values of  mixing angles are shown in Tab.~1. The results
in Tab.~1 show that a physically reasonable $\cos\beta$ can only
have a very small value with the above $\Sigma_v$ and $\Sigma_s$,
and this requires the Jaffe-Lipkin term dominance. The sea in the
baryon-meson state (\ref{BM}) only provides the orbital angular
momentum to the nucleon, and the Jaffe-Lipkin term (\ref{JL})
provides the negative polarized sea spin. Thus the importance of
the Jaffe-Lipkin term depends on the constraints of the sea quark
polarization of the nucleon.

\vspace{0.5cm}

\centerline{Table 1 The mixing angles }


\begin{footnotesize}
\begin{center}
\begin{tabular}{|c||c|}\hline
$~~~~~~~~\Sigma_s(\rm{II})=-0.69~~~~~~~~$&
$~~~~~~~~\Sigma_s(\rm{III})=-0.56~~~~~~~$\\ \hline
$~~\rm{sin}\alpha~~~$ \vline $~~~~\rm{sin}\beta~~~$ \vline
$~~~~\rm{sin}\theta~~$ & $~~\rm{sin}\alpha~~$ \vline
$~~~~\rm{sin}\beta~~~$ \vline $~~~\rm{sin}\theta~~~$
\\ \hline
$~\pm 0.200~$ \vline $~\pm 1.080~$ \vline $~~-0.408~$ & $~\pm
0.200$ \vline $~\pm 0.947~$ \vline $~~-0.452$
\\ \hline
$~\pm 0.400~$ \vline $~\pm 1.065~$ \vline $~~-0.429~$ & $~\pm
0.400$ \vline $~\pm 0.956~$ \vline $~~-0.436$
\\ \hline
$~\pm 0.600~$ \vline $~\pm 1.052~$ \vline $~~-0.452~$ & $~\pm
0.600$ \vline $~\pm 0.966~$ \vline $~~-0.419$
\\ \hline
$~\pm 0.800~$ \vline $~\pm 1.041~$ \vline $~~-0.472~$ & $~\pm
0.800$ \vline $~\pm 0.975~$ \vline $~~-0.405$
\\ \hline
\end{tabular}
\end{center}
\end{footnotesize}

\vspace{0.5cm}

From a strict sense, the sea spin $\Sigma_s$ has not been measured
directly, and also the Melosh-Wigner rotation factors should be
introduced into the so called spin term $\Sigma_v$ obtained from
hyperon and nucleon $\beta$-decays, and the flavor asymmetry and
SU(3) symmetry breaking should be important. Therefore the above
analysis needs to be updated. It would be more practical to
decompose the spin by the contributions from the quarks
$\Sigma_q=\Sigma_v+\frac{1}{2}\Sigma_s$, the antiquarks
$\Sigma_{\bar q}=\frac{1}{2}\Sigma_s$, and the orbital angular
momentum $\Lambda_s$, which still meet the condition
\begin{equation}
\Sigma_q+\Sigma_{\bar q}+\Lambda_s=1.
\end{equation}
The antiquark helicity distributions extracted from semi-inclusive
deep inelastic scattering experiments are consistent with zero
\cite{NSMCN}, in agreement with the small antiquark polarization
predicted in both the baryon-meson fluctuation model \cite{BM} and
a chiral quark model \cite{Che95}. There is still no direct
evidence for a large negative antiquark polarization in
experiments. We also point out here that there should be a
quark-antiquark asymmetry for the spin of the sea when flavor
decomposition is necessary \cite{BM}.

Since new measurements on the polarized structure functions for
both the proton and the neutron have become available, we will use
the measured $\Gamma^p$ and $\Gamma^n$ as inputs to study the
effects due to the inclusion of Jaffe-Lipkin and BM higher Fock
state terms in the nucleon wavefunction. Another aspect that we
need to take into account is that the $u$ and $d$ flavor
asymmetries should exist in both the valence and sea contents of
the nucleon. The observation of the Gottfried sum rule violation
in several processes \cite{NMC,GSRex} implies that there is an
important contribution coming from the lowest baryon-meson
fluctuation $p(uud D\bar D)=n(udD)\pi^+(u \bar D)$ of the proton
\cite{BM,SeaA}. This puts a constraint on the value of $\alpha$
for the BM mixing term. If one assumes an isospin symmetry between
the proton and neutron \cite{Ma92}, then the Gottfried sum rule
violation implies an asymmetry between the $u$ and $d$ sea
distributions inside the proton
\begin{equation}
\int_0^1 {\mathrm d} x \left[ \bar d(x) -\bar u(x) \right]=0.148
\pm 0.039.
\end{equation}
If we consider only the  $p(uud D\bar D)=n(udD)\pi^+(u \bar D)$
component inside the BM term and neglect flavor asymmetry in the
Jaffe-Lipkin term, then we get the constraint,
\begin{equation}
\sin^2 \alpha \cos^2 \beta=0.148.
\end{equation}
The $u$ and $d$ quark spins in the proton wavefunction should be
\begin{eqnarray}
\Delta u_{QM}=\cos^2\alpha \cos^2\beta \Delta u_0-
\frac{1}{3}\sin^2\alpha \cos^2 \beta \Delta d_0 + \sin^2 \beta
\Delta u_{JL};
\\
\Delta d_{QM}=\cos^2\alpha \cos^2\beta \Delta d_0-
\frac{1}{3}\sin^2\alpha \cos^2 \beta \Delta u_0 + \sin^2 \beta
\Delta d_{JL},
\end{eqnarray}
where $\Delta u_0=4/3$ and $\Delta d_0=-1/3$ are the $u$ and $d$
quark spins for the bare $qqq$ proton, and $\Delta u_{JL}$ and
$\Delta d_{JL}$ are the $u$ and $d$ quark spins for the
Jaffe-Lipkin term Eq.~(\ref{JL}) from $b$, $\varepsilon$, and $D$,
\begin{equation}
\Delta q_{JL}=(1-\frac{4}{3} \sin^2\theta) \Delta
q_0+\frac{1}{4}\Sigma_s
\end{equation}
for $q=u, d$ in case of only charge neutral $Q\bar Q$'s with $u$
and $d$ flavors. Substituting the above $\Delta u_{QM}$ and
$\Delta d_{QM}$ into Eqs.~(\ref{vMu}) and (\ref{vMd}), we get
\begin{equation}
\langle M_u \rangle=0.598, ~~\langle M_d \rangle=0.878,
~~~{\mathrm and}~~~ r_{d/u}=\langle M_d \rangle/\langle M_u
\rangle=1.47, \label{Mudr}
\end{equation}
for $\beta=0$ without the Jaffe-Lipkin term. We find that the $u$
and $d$ flavor asymmetry $r_{d/u}$ is reduced compared to
Eq.~(\ref{Mud}) and this shows that the $p(uud D\bar
D)=n(udD)\pi^+(u \bar D)$ fluctuation produces a more reasonable
$d/u$ Melosh-Wigner rotation asymmetry than in the naive picture
with the bare nucleon state of only three valence quarks
\cite{Ma96}.

In fact, we should also include other baryon-meson fluctuations in
a more realistic picture of intrinsic sea quarks \cite{BM}, such
as $p(uud U \bar U)=\Delta ^{++}(uuU)\pi^-(d \bar U)$ for the
intrinsic $U\bar U$ quark-antiquark pairs and $p(uud S \bar S)=
\Lambda (ud S)K^+(u \bar S)$ for the intrinsic strange
quark-antiquark pairs. In this case we can write the baryon-meson
term as
\begin{eqnarray}
\sin\alpha \cos \beta \left |[BM]^\uparrow \right> =
A(n\pi^+)\left|n\pi^+\right>+A(\Lambda K^+)\left|\Lambda
K^+\right>+ A(\Delta^{++}\pi^-)\left|\Delta^{++}\pi^-\right>,
\end{eqnarray}
where we take the baryon-meson configuration probabilities
$P(p=BM)=[A(BM)]^2$ as
\begin{equation}
P(p=n\pi^+)\sim 15\%; ~P(p=\Lambda K^+) \sim
3\%;~P(p=\Delta^{++}\pi^-)\sim 1\%,
\end{equation}
as estimated from a reasonable physical picture \cite{BM}. With
the above baryon-meson fluctuations considered, we find,
\begin{equation}
\langle M_u \rangle=0.624, ~~\langle M_d \rangle=0.912,
~~~{\mathrm and}~~~ r_{d/u}=\langle M_d \rangle/\langle M_u
\rangle=1.46, \label{Mudr2}
\end{equation}
which are close to Eq.~(\ref{Mudr}), the case with only $p=n\pi^+$
fluctuation. Thus our above analysis supports a reasonable picture
of a dominant valence three quark component with a certain amount
of the energetically-favored baryon-meson fluctuations \cite{BM},
as a ``minimal" quark model for the spin relevant observations in
DIS processes and also for several phenomenological anomalies
related to the flavor content of nucleons \cite{BM}. Of course, we
can also include the necessary other higher $5q$ Fock states
approximated in terms of the BM state and the Jaffe-Lipkin state.

The gluon distribution of a hadron is usually assumed to be
generated from the QCD evolution. However, it has been pointed in
Ref.~\cite{BS} that there exist intrinsic gluons in the
bound-state wavefunction. Therefore we could also consider the 
possibility of including a $(qqqg)$ Fock state in our 
description. Unfortunately the gluon is always a relativistic 
particle, and it is not easy to incorporate it in the present 
framework. We must use a relativistic approach from the start, 
such as the one given in Ref.~\cite{Bro00}.

\section{Summary and discussion}

We investigated the spin structure of the nucleon in a simple
quark model. First, we studied the possible effect due to the
Melosh-Wigner rotation. We found that an introduction of an
up-down quark flavor asymmetry in the Melosh-Wigner rotation
factors can reproduce the present experimental data of the
integrated spin structure functions for both the proton and the
neutron within a simple SU(6) quark model with only three valence
quarks. And then, we discussed the contributions to the nucleon
spin from various components of the nucleon wavefunction. The
calculated results indicate that the baryon-meson state of
Jaffe-Lipkin with vector meson is important concerning the sea
polarization of the nucleon regardless of the existence of states
which include the pseudoscalar mesons.

The Melosh-Wigner rotation is one of the most important
ingredients of the light-cone formalism. Its effect is of
fundamental importance in the spin content of hadrons, and it is
mainly due to the transverse momentum of quarks in the nucleon.
Actually, it reflects some relativistic effects of a quark system.
On the other hand, the simple quark model discussed here includes
the baryon-meson fluctuation component in the nucleon
wavefunction, which is a nonperturbative effect. The present
investigation shows that relativistic and nonperturbative effects
are very important in order to understand the spin structure of
the nucleon. In the simple quark model, the bare three quark
component and the baryon-meson state with a pesudoscalar meson,
are still important concerning the proton spin problem in
polarized structure functions, after we take into account the
Melosh-Wigner rotation effect.

{\bf ACKNOWLEDGMENTS: } This work is partially supported by
National Natural Science Foundation of China under Grant Numbers
19975052 and 19875024, No.~19775051, and by Fondecyt (Chile)
postdoctoral fellowship 3990048, by Fondecyt (Chile) grant 1990806
and by a C\'atedra Presidencial (Chile).

\end{document}